\documentclass[doublecol]{epl2}
\usepackage{color}
\usepackage{epsfig}
\usepackage{graphicx}
\usepackage{dcolumn}
\usepackage{bm}
\title{The unified quantum wave equation}
\shorttitle{The unified quantum wave equation}
\author{Arbab I. Arbab\inst{}\footnote{aiarbab@uofk.edu; arbab.ibrahim@gmail.com}}

\institute{
  \inst{} Department of Physics,
Faculty of Science, University of Khartoum, P.O. Box 321, Khartoum
11115, Sudan
}

\abstract{
 The quaterionic formulation of quantum mechanics yields the  unified quantum wave equation (UQWEs). From these equations, Dirac, Klein - Gordon and Schrodinger equations can be derived.  While the UQWEs represent a matter wave (de Broglie), the Maxwell equations represent a transverse wave (field). Owing to UQWEs,  the spin-0 and spin-1/2 particle are described by a wavepacket consisting of waves traveling to the left and to the right with speed of light. UQWEs show that spin-0 and spin-1/2 are in continuous states of creation and annihilation that are compatible with Heisenberg uncertainty relation. The creation - annihilation process is a result of the time translation property of the particle wavefunction. These are $E'=E-im_0c^2$ and $E'=E\pm m_0c^2$,  for Klein-Gordon' and Dirac' particles, respectively. It is found that $\frac{\hbar}{m_0c^2}$ is the period of the creation -annihilation process.}

\begin{document}

\maketitle

\section{Introduction}

Quantum mechanics has been developed by Schrodinger who employed de Broglie hypothesis about the matter wave. The solution of Schrodinger is a plane wave. However, a particle should be represented by a collection of waves (wavepacket). Schrodinger equation is valid for a particle moving at non-relativistic speed. Later, Klein and Gordon developed a quantum theory based on Einstein's relativity. The resulting equation represents the motion of spin -0 massive particles. However, the probability density obtained from this equation is not positive definite. This urges physicists to replace it by a more convenient equation. For this Dirac developed his quantum theory. To remedy this impasse, Dirac presented a first order differential equation in space and time.

Maxwell had used the quaternions formulation to write his electromagnetic equations. The resulting number of equations are too many (20 equations). These equations look absurd. Only after Gibbs and Heaviside invented the vector analysis, Maxwell equations, in their present form, became conspicuous.
Since that time quaternions had been absent from physics except for some recent limited trials.
The surprising work has come recently when I reintroduce  quaternions in quantum mechanics employing new ideas ~\cite{arb}. As a result,  unified quantum wave equation is obtained. From this equation we have derived the Dirac, Klein-Gordon and Schrodinger equations ~\cite{arb1}. The UQWEs express Dirac's equation as a second-order wave equation ~\cite{arb, arb3}. Consequently, Klein-Gordon as well as Dirac equations are of the same mathematical structure. With the aid of Arbab-Widatallah complex transformations, Dirac equation in its new form is derived from the UQWEs using the complex mass transformation $m_0\rightarrow i\,\beta \,m_0$ ~\cite{wida, hooft}. Moreover, Dirac and Klein-Gordon equations are found to stem from a massless wave equation upon making the mass translations (shift/rotation). Like Maxwell equations, UQWEs involve scalar and vector waves. The scalar wave represents longitudinal wave while vectorial wave represents transverse waves. The vectorial wave can be some sort of spin (or polarization) wave that some magnetic  systems have exhibited. It can also be related to the field associated with the particle (electric, magnetic, gravitational,... etc). In a recent work, we have shown that for spin-0 particles, the vector field is related to its acceleration~\cite{arbi}. In our present case, one finds $(\varphi\Leftrightarrow \psi_0$) and ($\vec{A}\Leftrightarrow\vec{\psi}$) as fundamental fields. Aharonov and Bohm demonstrated that the real fields that the quantum nature of the particle reveals are the vector potential $\vec{A}$  and scalar potential $\varphi$  ~\cite{ahar}. The phenomenon in which an electron is affected by $\vec{A}$ and $\varphi$ produces the interference pattern of wavefunction is called Aharonov-Bohm effect. This is confirmed experimentally by setting up an environment with $\vec{A}$ and $\varphi$, while having zero electric and magnetic fields. This poses the question whether  $\vec{A}$ and $\varphi$ are more fundamental than $\vec{E}$ and $\vec{B}$. Moreover,  $\vec{A}$ can be decomposed ($\vec{A}_\bot$) in transverse and parallel ($\vec{A}_\|$) components. In 1930 Fermi showed that $\vec{A}_\|$ and $\varphi$ give rise to the instantaneous Coulomb interactions between the charged particles, whereas $\vec{A}_\bot$ accounts for the electromagnetic radiation of charged moving particles.
\section{Universal quantum wave equation}
We have recently derived a system of unified quantum wave equations ~\cite{arb, arb1}
\begin{equation}
\vec{\nabla}\cdot\vec{\psi}-\frac{1}{c^2}\frac{\partial \psi_0}{\partial t}-\frac{m_0}{\hbar}\,\psi_0=0\,,
\end{equation}
\begin{equation}
\vec{\nabla}\psi_0-\frac{\partial \vec{\psi}}{\partial t}-\frac{m_0c^2}{\hbar }\,\vec{\psi}=0\,,
\end{equation}
and
\begin{equation}
\vec{\nabla}\times\vec{\psi}=0\,.
\end{equation}
Equations (1) - (3) can be solved to give
\begin{equation}
\frac{1}{c^2}\frac{\partial^2\psi_0}{\partial t^2}-\nabla^2\psi_0+2\left(\frac{m_0}{\hbar}\right)\frac{\partial\psi_0}{\partial t}+\left(\frac{m_0c}{\hbar}\right)^2\psi_0=0\,,
\end{equation}
and
\begin{equation}
\frac{1}{c^2}\frac{\partial^2\vec{\psi}}{\partial t^2}-\nabla^2\vec{\psi}+2\left(\frac{m_0}{\hbar}\right)\frac{\partial\vec{\psi}}{\partial t}+\left(\frac{m_0c}{\hbar}\right)^2\vec{\psi}=0\,.
\end{equation}
This is a dissipative wave equation for spin - 0 particle. It is a generic Telegraphy equation representing  signal transmission. Such a wave arises when friction or other dissipative force produces a damping (proportional to the velocity of vibration), whose effect in the wave equation is the inclusion of the term proportional to $\frac{\partial\psi}{\partial t}$.

The solution of  eq.(4)/or eq.(5) is of the form
$$
\psi_0(x, t)=A\exp(-\frac{m_0c^2}{\hbar}t)\exp(\pm 2\pi\,i(ckt-\vec{k}\cdot \vec{r}))\,,\,\,\, $$
where $A=\rm const.$ and $\vec{k}$ is the propagation constant. It represents an undistorted damped wavepacket moving to the left and right.

It is interesting to note that eqs.(4) and (5) represent a scalar wave (longitudinal) and a vector wave that are concomitant with the particle motion. The vector wave is only a feature of our present equation and does not exist in Schrodinger, Dirac or Klein-Gordon description. Therefore, other  physical properties can be associated with this vector nature of the particle. Hence, the complete physical description of the particle will be performed in terms of these two waves.

Let us now write
\begin{equation}
\psi_0(r,t)=\exp\, (-\frac{m_0c^2}{\hbar}\, t)\, \varphi(r,t)\,.
\end{equation}
Substitute eq.(6) in eq.(4) to get
\begin{equation}
\frac{1}{c^2}\frac{\partial^2\varphi}{\partial t^2}-\nabla^2\varphi=0\,.
\end{equation}
Hence, $\varphi(r, t)$ satisfies the wave equation. Therefore, the UQWE is a relativistic equation. It is remarkable that the UQWEs describe the particle by a scalar and vector quantities. Thus, the full description of the particle motion can be made using these two quantities only.
Moreover, since eqs.(5) \& (6) are of a Telegraph-type equation that represents the motion of the electric signal in a wire, then the motion of a particle in space mimics  signal propagation. Moreover, eq.(5 and (6) are of special nature that describes the propagation of an undistorted signal along the wire. Hence, the particle of spin-0 travels in space undistortedly.

Dirac's equation can be written as ~\cite{dirac}
\begin{equation}\label{1}
\frac{1}{c}\frac{\partial\psi}{\partial t}+\vec{\alpha}\cdot\vec{\nabla}\psi+\frac{im_0c\,\beta}{\hbar}\,\psi=0\,.
\end{equation}
where   $\beta=\left (\begin{array}{cc}
  1 & 0 \\
  0 & -1 \\
\end{array}\right),$    $\alpha=\left (\begin{array}{cc}
 0 & \vec{\sigma}   \\
   \vec{\sigma}  & 0 \\
\end{array}\right),$ $\alpha^2=\beta^2=1$ and $\vec{\sigma}$ are the Pauli matrices.
Equation (8) can be written as
\begin{equation}\label{1}
\frac{1}{c}\frac{\partial\psi}{\partial t}+\frac{i\,m_0c\,\beta}{\hbar}\psi=-\vec{\alpha}\cdot\vec{\nabla}\psi\,.
\end{equation}
Squaring the two sides of eq.(9) yields
\begin{equation}\label{1}
\left(\frac{1}{c}\frac{\partial}{\partial t}+\frac{i\,m_0c\,\beta}{\hbar}\right)^2\psi=\left(-\vec{\alpha}\cdot\vec{\nabla}\right)^2\psi\,.
\end{equation}
Since $\alpha^2=\beta^2=1$, eq.(10) yields
\begin{equation}
\frac{1}{c^2}\frac{\partial^2\psi}{\partial t^2}-\nabla^2\psi+2\left(\frac{i\,m_0\beta}{\hbar}\right)\frac{\partial\psi}{\partial t}-\left(\frac{m_0c}{\hbar}\right)^2\psi=0\,.
\end{equation}
It is remarkable to know that eq.(11) can be obtained directly from eq.(4) if we let
\begin{equation}
m_0\rightarrow i\beta\, m_0\,,
\end{equation}
in eq.(4). This entitles the wavefunction $\psi$ to be represented by a two-component construct, viz., $\psi=\left (\begin{array}{c}
  \psi_+\\
  \psi_- \\
\end{array}\right)$\,.

Let us now write
\begin{equation}
\psi(r,t)=\exp\, (-\frac{i\,m_0c^2\beta}{\hbar}\, t)\, \chi(r,t)\,.
\end{equation}
Substituting eq.(13) in eq.(11) yields
\begin{equation}
\frac{1}{c^2}\frac{\partial^2\chi}{\partial t^2}-\nabla^2\chi=0\,.
\end{equation}
Once again, under the energy translation of the particle's wavefunction both Dirac and Klein-Gordon equations describe a massless particle.
This is the equation of a massless particle (wave).

Let us now write the wavefunction
\begin{equation}
\psi_0(r, t)=\exp\,(-\vec{\kappa}\cdot\,\vec{r})\,\varphi(r,t)\,,\qquad \kappa=\rm const.\,,
\end{equation}
 and substitute it in eq.(4) to get
 \begin{equation}
\frac{1}{c^2}\left(\frac{\partial}{\partial t}+\frac{m_0c^2}{\hbar}\right)^2\varphi-\left(\vec{\nabla}-\vec{\kappa}\right)^2\varphi=0\,,
\end{equation}
or
\begin{equation}
\frac{1}{c^2}\frac{\partial^2\varphi}{\partial \eta^2}-\nabla\,'^{\,2}\varphi=0\,,
\end{equation}
where
\begin{equation}
 \frac{\partial}{\partial \eta}= \frac{\partial}{\partial t}+\frac{m_0c^2}{\hbar}\,,\qquad \vec{\nabla}\,'=\vec{\nabla}-\vec{\kappa}\,.
\end{equation}
Equations (17) and (18) are the wave equation for a massless particle interacting with an external field described by $\kappa$. This can be compared with a massless particle interacting with electromagnetic potentials $\vec{A}$ and $\varphi$ where $\vec{A}\equiv -\frac{i\hbar}{e}\,\vec{\kappa}$ and $V\equiv i\, m_0c^2$.

Similarly, let us now write the wavefunction
\begin{equation}
\psi(r, t)=\exp \,(-\vec{\kappa}\cdot\,\vec{r})\,\chi(r,t)\,,\qquad \kappa=\rm const.\,,
\end{equation}
and substitute it in eq.(11) to get
\begin{equation}
\frac{1}{c^2}\left(\frac{\partial}{\partial t}+i\frac{m_0c^2}{\hbar}\right)^2\chi-\left(\vec{\nabla}-\vec{\kappa}\right)^2\chi=0\,,
\end{equation}
or
\begin{equation}
\frac{1}{c^2}\frac{\partial^2\chi}{\partial \tau^2}-\nabla\,'^{\,2}\chi=0\,,
\end{equation}
where
\begin{equation}
 \frac{\partial}{\partial \tau}= \frac{\partial}{\partial t}+i\frac{m_0c^2}{\hbar}\,,\qquad \vec{\nabla}\,'=\vec{\nabla}-\vec{\kappa}\,.
\end{equation}
Equation (22) can be seen as representing a massless particle interacting with an external vector field $\vec{\kappa}$ in potential energy $m_0c^2$. Thus, Dirac and Klein-Gordon particles, with mass $m_0$, are equivalent to  massless  particles (waves) interacting, with  constant real and imaginary scalar potential with the same constant vector potential, respectively. These correspond to $E'_{KG}=E+im_0c^2$, $E'_{D}=E+m_0c^2$, and $\vec{p}\,\,'=\vec{p}+i\,\hbar\vec{\kappa}$, for Klein-Gordon (KG) and Dirac (D) particles, respectively.

Consequently, eqs.(15) and (19) can be seen as representing the local gauge transformation of the Klein-Gordon and Dirac wavefuncions, $\chi$ and $\varphi$, respectively. In this case the four vector potential $A_\mu=(A_0\,,\vec{A})$ will be $(m_0c^2\,,-\frac{i\hbar}{e}\,\vec{\kappa})$ for Dirac and $(im_0c^2\,,-\frac{i\hbar}{e}\,\vec{\kappa})$ for Klein-Gordon. In electromagnetism, the gauge transformation is obtained via $A_\mu'=A_\mu+\partial_\mu \lambda$. This corresponds, in our present theory, to $\lambda=-\frac{i\hbar}{e} \vec{\kappa}\cdot\vec{r}$. This analogy is quite interesting. The terms $A_\mu$  and $\partial_\mu \lambda$ can be seen as describing the transverse an paralel components of the vector potential, respectively. We have shown recently that the parallel component of $\vec{A}$, i.e., $\partial_\mu \lambda$, gives rise to a longitudinal wave ~\cite{arbi}.

The dispersion relation arising from eq.(4) is
\begin{equation}
\omega_\pm=\frac{m_0c^2}{\hbar}i\pm ck\,,
\end{equation}
so that the group velocity is
\begin{equation}
v_g=\frac{\partial \omega}{\partial k}=\pm c\,.
\end{equation}
\section{Creation and annihilation of particles}
It interesting that the mass of the particle disappears in eq.(14) while appears in eq.(11). Hence,  at a time $\tau$ (after an interval of $\frac{\hbar}{m_0c^2}$ the particle loses its mass (annihilates) and then again being created after the same time. Therefore, as time goes on the particle experiences continuously a processes of creation and annihilation. This process is governed by the time of uncertainty owing to the Heisenberg uncertainty relation ($\Delta t\, \Delta E\ge \hbar$).

Under the transformation
\begin{equation}
\frac{\partial}{\partial \tau}=\frac{\partial}{\partial t}+i\frac{m_0c^2}{\hbar}\,\beta\,,
\end{equation}
eq.(11) can be written as
\begin{equation}
\frac{1}{c^2}\frac{\partial^2\psi}{\partial \tau^2}-\nabla^2\psi=0\,.
\end{equation}
And under the transformation
\begin{equation}
\frac{\partial}{\partial \eta}=\frac{\partial}{\partial t}+\frac{m_0c^2}{\hbar}\,,
\end{equation}
eq.(4) becomes
\begin{equation}
\frac{1}{c^2}\frac{\partial^2\psi}{\partial \eta^2}-\nabla^2\psi=0\,.
\end{equation}
Using eq.(8), Dirac equation, eq.(11), can be written as
\begin{equation}
\frac{1}{c^2}\frac{\partial^2\psi}{\partial t^2}-\nabla^2\psi-2\left(\frac{m_0c\,i}{\hbar}\right)\beta\vec{\alpha}\cdot\vec{\nabla}\psi+\left(\frac{m_0c}{\hbar}\right)^2\psi=0\,,
\end{equation}
which under the transformation
\begin{equation}
\vec{\nabla}\,'=\vec{\nabla}+i\frac{m_0c\beta}{\hbar}\,\vec{\alpha}\,,
\end{equation}
becomes
\begin{equation}
\frac{1}{c^2}\frac{\partial^2\psi}{\partial t2}-\nabla'^2\psi=0\,.
\end{equation}
Once again, this is a wave equation of a massless particle. Thus, a particle of spin-0 or spin-1/2 undergoes a process of creation and annihilation during its propagation in space and time.

Hence, eqs.(15) \& (16) and eqs.(17) \& (18) are compatible with eqs.(6) \& (14). Therefore, while Dirac's particle undergoes a virtual process of creation and annihilation, Klein-Gordon's particle undergoes the same process in real time.
It is interesting to observe that eqs.(7) and (18) are connected by the energy translation, viz., $E\,'=E-i\,m_0c^2$ and $E\,'=E\pm m_0c^2$, to eqs.(6) and (13), respectively. Hence, a massless wave equation  can be obtained from Dirac equation by employing the energy translation (shift), $E\,'=E\pm m_0c^2$, instead of setting the particle mass to zero, and vice versa. These may be attributed to time advance and time retard translation. Similarly, Klein-Gordon equation can be obtained from UQWE by employing the energy translation (rotation), $E\,'=E- i\,m_0c^2$ . This may be attributed to time rotation of the wavefunction. Imaginary mass is like imaginary frequency, designates a dissipation in the oscillating system.
Moreover, these transformations are equivalent to set, $\omega\,'=\omega-i\,\omega_c$\,, and $\omega\,'=\omega\pm\,\omega_c$\,, where $\omega_c=\frac{m_0c^2}{\hbar}$.

The wavefunction of Dirac particle is a wavepacket consisting of waves traveling to the right and left with speed of light in opposite direction. The dimension of this wavepacket is $L=\frac{\hbar}{m_0c}$. The transformation in eq.(30) tells us that the creation and annihilation processes occur periodically over space and time. One can argue that the spin of the Dirac particle is due to the rotation of the two waves comprising the particle around each other. For an electron, each wave (identity) has a mass half that of the electron, i.e., $m_0/2$. The spin angular momentum will thus be
\begin{equation}
S=I\omega=\Sigma_i m_ir_i^2\omega_i=m_1r_1^2\omega_1+m_2r_2^2\omega_2\,,\omega_1=\omega_2=\frac{c}{r}\,.
\end{equation}
Hence,
\begin{equation}
S=\frac{1}{2}\,\hbar\,\,,\qquad r=L/2.
\end{equation}
This coincides with the quantum value predicted by Dirac. It has long been believed that the spin is not a classical concept!
It thus becomes obvious that the electron spin is analogous to the angular momentum of the classical circularly polarized wave. It has been shown by  Belinfante in 1939, that it is possible that the electron spin can be considered as an angular momentum, generated by the circulation of the energy flow in the field of electron wave ~\cite{spin}. It is of importance to remark that the property in eq.(30) is not applicable to eq.(4) of spin -0 particles.

It is interesting to notice that because of the damping nature of the the spin-0 zero particles (eq.(6)), their interactions are of short range. This is in agreement with Yukawa's theory. This is unlike the interactions of the spin -1/2 particles which  have oscillatory wave nature, as evident from eq.(14). Owing to this property, spin-1/2 particles have long range interactions. The conversion of mass into energy and ice versa are a manifestation of Einstein's mass-energy equivalence, since both equations emerge from this equation.
\section{Conclusions}
We have developed  unified  quantum mechanics wave equations that have an analogy with Maxwell equations, and yield Dirac and Klein - Gordon equations. These unified equations represent, like Maxwell equations, scalar (longitudinal) and vectorial (transverse) waves.
In the context of these equations, the solution of the  modified Klein-Gordon equation  is that of the normal Klein-Gordon equation where the frequency will be $\omega\,'=\omega-i\,\omega_c$. However, the solution of the Dirac equation is obtained from the wavefunction  of the standard wave equation by allowing the frequency of the wave to be $\omega\,'=\omega\pm\,\omega_c$.  Moreover, the Dirac and Klein-Gordon equations are found to be equivalent to an equation of massless particle interacting with constant vector and scalar potentials. The transformed wavefunction for Dirac and Klein-Gordon equation are that of the  local gauge transformation with a linear gauge function. Dirac and Klein-Gordon equations are shown to exhibit creation - annihilation process consistent with Heisenberg uncertainty equation.
\acknowledgments

\end{document}